# A Modified DTC with Capability of Regenerative Braking Energy in BLDC driven Electric Vehicles Using Adaptive Control Theory

Shiva Geraee[a,*], Hamed Mohammadbagherpoor[b], Mehdi Shafiei[c], Majid Valizadeh[d], Farshad Montazeri[e], Mohammad Reza Feyzi[a]

[a]Faculty of Electrical and Computer Engineering, University of Tabriz, Iran

[b]Faculty of Electrical and Computer Engineering, North Carolina State University, USA

[c]Department of Electrical and Electronic Engineering and Computer Science, Queensland University of Technology, Australia

[d]Department of Electrical and Electronic Engineering, University of Ilam, Iran

[e]Center of Excellence for Power System Automation and Operation Electronic Engineering, Iran University of science and Technology, Iran

**Abstract**

This paper represents a novel regenerative braking approach for the Electric Vehicles. The proposed method solves the short-range problem which is corresponding to the charge of the battery pack. The DTC switching algorithm has been modified to recover the electrical energy from Electrical Vehicle (EV), driven by Brushless DC motor, without using the additional power converter or the other electrical energy storage devices. During regenerative braking process, different switching pattern is applied to the inverter to convert the mechanical energy to the electrical energy through the reverse diodes. This switching pattern is different from the normal operation due to the special arrangement of voltage vectors which is considered to convert the mechanical energy to electrical energy. The state of charge of the battery is used as a performance indicator of the method. Simultaneously, a model reference adaptive system has been designed to tune the system's parameters. Several simulations are carried out to validate the performance and effectiveness of the proposed methods. The results show the high capability and performance of the designed method.

Keywords: Electric Vehicle, Regenerative Braking, Brushless DC Motor, Direct Torque Control, Model Reference Adaptive System.

**Introduction**

Nowadays, due to the concern of global warming, fossil fuels cost and uncertainty of oil supply, the necessity of using renewable energies become more obvious [1-3]. Therefore, the government's policies have been changed to encourage automobile manufactures to allocate research budget for EVs [1,4]. In compare with the conventional vehicles there is still some drawbacks of EVs that can be noted such as, battery pack, charging system and a short driving mileage due to battery charging capacity [4-6]. Therefore, under government's instruction and the people demands, automobile manufacturers have tried to improve the products in terms of standard quality and fuel efficiency [7].

In order to increase the efficiency of the EVs, high-tech equipment such as sensors, extra storage devices, and inverter circuits can be employed. However, these technologies on one hand may make EVs complicated, and on the other hand increase the total cost of producing EVs [23]. Therefore, researchers are trying to tackle this problem by modifying the regenerative EVs' braking system. The battery pack, the motor drive system, and the converter controller are three significant components of EVs [5]. Motor drive system technology plays a crucial role in power transfer mechanism [3]. The common electric motors that are used in EVs are switched reluctance machine (SRM), induction motor (IM) and brushless direct-current (BLDC) motor [3]. Among them the BLDC motors are the preferred one. The most applications of the Brushless DC motors can be categorized into the servo drives, appliances, medical equipment and a broad spectrum of power systems [8,9]. Its popularity goes back to the great efficiency, suitable torque–speed characteristic, dependability, stableness, lower noise and simple structure to control [9-11]. However, the significant defect of BLDC motors is related to the torque ripple which is resulted to cogging torque [12].

Also, awareness of the exact position of the rotor is one of the key issues in this kind of motor to sustain the windings in the right direction of rotation [13]. These positions are sensed by the Hall sensors owing to their cost-effectiveness, especially where the phase currents should be commutated on and off [14, 15]. The proximity sensors detect the pole's sign of the rotor when the rotor magnetic poles are passing in front of the sensors [15]. Assessing the received signals determines the precise sequence of commutation. The signals can be decoded by understanding the combinational logic. The outputs are 6 voltage vectors. Therefore, the firing commands are applied to the phases to conduct them to 120 electrical degrees.

Various methods have been proposed to control BLDC motor such as dc link current control, direct torque control, hysteresis current control and pulse width modulator control [16,17], that among them Direct Torque Control (DTC) is more reliable [18]. A three-phase switching circuit is used to control the motors based on the received command from the controller [3]. Switching pattern in DTC can be altered in diverse ways which differ in purpose. Heretofore, a multitude of articles has published to improve the regenerative braking approaches of EVs in various procedures in distinct motors [19-26]. The energy efficiency is a crucial attribute [27], and reusing the brake energy can make a profound contribution, a distinct difference between conventional vehicles and EVs. By regenerative braking the energy is reversed into the battery within braking so that torque is used for braking the motor, and the motor generates power consequently [13,15]. Recovering potential and kinetic energy during declaration or stopping can increase the range of the car's path. Adding equipment such as sensors, extra storage devices, and new inverter circuits, in some of the proposed methods only make system complicated [23]. In the proposed method, it has been attempted to recover the kinetic energy only by changing the switching pattern without the major revolution in structure. Voltage vectors are sorted based on the roles that they play in reducing or augmenting in torque and flux. The controller is the heart of the system for both driving and regenerative braking of motors. A very typical control system is proportional-integral (PI) controller, which is popular due to a simple structure and high torque and speed responses. Nonetheless, the performance of this controller is not acceptable in the presence of speed changes, parameters fluctuation and load effect [8,28]. In this paper, a Model Reference Adaptive System (MRAS) is applied for tuning the controller parameters automatically and track the reference speed signal.

**Dynamic model of Electric Vehicle**

The motion and acceleration of the vehicle can be analyzed by the forces applied to it. In the motoring mode, resistance forces acting on EV as shown in Figure 1, can be equivalent to the following formulas [5]:

$$F_{tot} = F_{rr} + F_{hc} + F_{ad} + F_{la} \tag{1}$$

$$F_{rr} = M.g.f \quad \text{is the rolling resistance} \tag{2}$$

$$F_{hc} = M.g.\sin\psi \quad \text{is the hill climbing force} \tag{3}$$

$$F_{ad} = \frac{1}{2}\rho.C_d.A.v^2 \quad \text{is the aerodynamic drag} \tag{4}$$

$$F_{la} = M \cdot \frac{dv}{dt} \qquad \text{is the acceleration force} \tag{5}$$

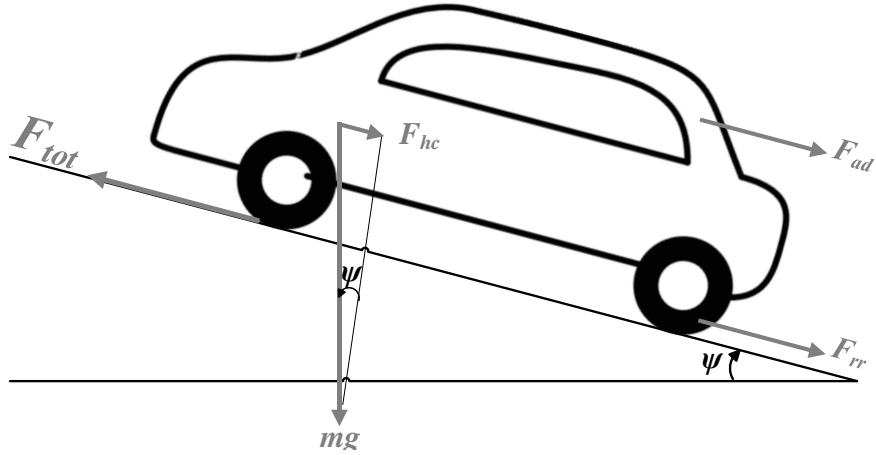

Figure. 1. The forces acting on a vehicle moving along slope

Where $g$ is the acceleration of gravity, $f$ is the coefficient of rolling resistance, $\rho$ is the density of the ambient air, $C_d$ is aerodynamic drag coefficient, $v$ is vehicle speed, $A$ is vehicle frontal area, $M$ is the overall mass of the vehicle, $\psi$ is the slope angle.

The total torque load can be calculated as (6) in:

$$T_L = F_{tot} * R_w \tag{6}$$

where $R_w$ stands for the radius of the tire.

An electric motor is responsible to prevail over load force by producing tractive force [29]. BLDC motors are the common motors that are used in EVs to increase their performances. A general switching circuit of the BLDC drive system has been shown in Figure 2. The equivalent dynamic model of BLDC is represented as equation (7) [18],

$$\begin{bmatrix} V_{an} \\ V_{bn} \\ V_{cn} \end{bmatrix} = \begin{bmatrix} R & 0 & 0 \\ 0 & R & 0 \\ 0 & 0 & R \end{bmatrix} \begin{bmatrix} i_a \\ i_b \\ i_c \end{bmatrix} + \begin{bmatrix} L & 0 & 0 \\ 0 & L & 0 \\ 0 & 0 & L \end{bmatrix} \frac{d}{dt} \begin{bmatrix} i_a \\ i_b \\ i_c \end{bmatrix} + \begin{bmatrix} e_a \\ e_b \\ e_c \end{bmatrix} \qquad (7)$$

$$L = L_s - L_m \qquad (8)$$

where $V_{an}$, $V_{bn}$, and $V_{cn}$ are the phase voltages, $i_a$, $i_b$, and $i_c$ are the phase currents, $R$ is the phase resistances, $L$ is the equivalent inductance in each stator windings, $Ls$ are inductance of each phase, $Lm$ isthe mutual inductance between phases, and $e_a$, $e_b$, and $e_c$ are back-EMFs of each phase [8].

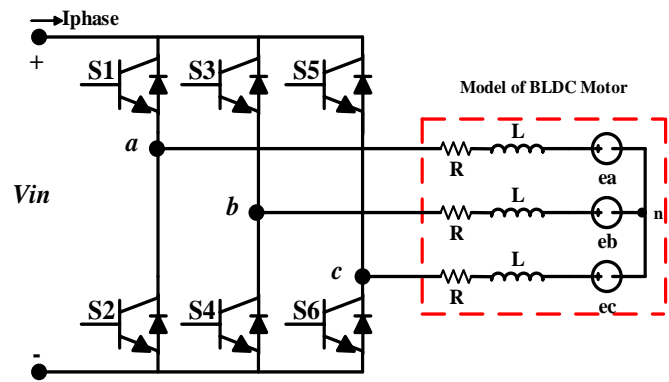

Figure. 2. A general schematic of BLDC drive

The ideal current and back-EMF signals for one of the phases have been illustrated in Figure 3.

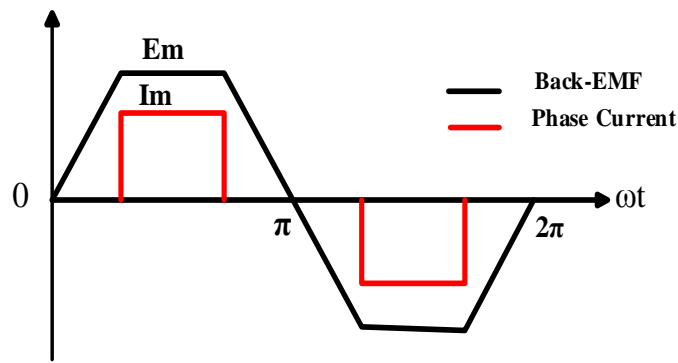

Figure. 3. Ideal waveform of back-EMF and curren

**Direct torque control of BLDC**

Direct Torque Control (DTC) was proposed by Takahashi and Noguchi in 1986 and Depenbrock in 1988, which was implemented on induction motors [30]. Due to its high performance in controlling the variable speed drivers, it is prevalent among the industrial applications. This method controls the flux linkage and electromagnetic torque simultaneously and independently. Also, DTC is robust against uncertain parameters, swift torque response, simplicity and high dynamic performance. However, it suffers from the torque ripples and fluctuating frequencies [15,18].

The electromagnetic torque plays an important role in DTC method. Equation (9) (9) describes the electromagnetic torque in stationary ($\alpha\beta$) reference frame [30]:

$$T_{em} = \frac{3}{2}\frac{P}{2}[\frac{d\varphi_{r\alpha}}{d\theta_e}i_{s\alpha} + \frac{d\varphi_{r\beta}}{d\theta_e}i_{s\beta}] = \frac{3}{2}\frac{P}{2}[\frac{e_\beta}{\omega_e}i_{s\alpha} + \frac{e_\alpha}{\omega_e}i_{s\beta}]$$

where $\omega_e$ are the electrical rotor position, $i_{s\alpha}$, $i_{s\beta}$ are stator currents and $e_\alpha$, $e_\beta$ are back- EMF respectively, $\theta_e$ is the electrical rotor position and $\varphi_{r\alpha}$ and $\varphi_{r\beta}$ are reference frame electromagnetic torque.

DTC operates based on the set of specific instructions to decide on the right voltage vector between six or eight space vectors. Table 1 shows the possible space vectors and patterns that DTC uses to choose the right voltage. Commands are chosen based on the hysteresis controller outputs. The difference between estimated and real values of torque and flux are the inputs of hysteresis controller. This error between real and estimated amount for flux linkage is named $F_{ST}$ and for electromagnetic torque $T_{ST}$. "TI" expression indicates the demand for increasing the torque since the actual extent torque is lower than the estimated extent."TD" expression shows that less electromagnetic torque is required. The "FI" and "FD" expressions are the same as torque which indicates the demand for increasing or decreasing of the flux, respectively. Furthermore, "F" expression is used to show the unchanged cases [31].

| $F_{st}$ | $T_{st}$ | Sector | | | | | |
|---|---|---|---|---|---|---|---|
| | | Θ1 | Θ2 | Θ3 | Θ4 | Θ5 | Θ6 |
| FI | TI | V1(100001) | V2(001001) | V3(011000) | V4(010010) | V5(000110) | V6(100100) |
| | TD | V6(100100) | V1(100001) | V2(001001) | V3(011000) | V4(010010) | V5(000110) |
| | TI | V2(001001) | V3(011000) | V4(010010) | V5(000110) | V6(100100) | V1(100001) |
| | TD | V1(100001) | V2(001001) | V3(011000) | V4(010010) | V5(000110) | V6(100100) |
| FD | TI | V0(010101) | V0(010101) | V0(010101) | V0(010101) | V0(010101) | V0(010101) |
| | TD | V2(001001) | V3(011000) | V4(010010) | V5(000110) | V6(100100) | V1(100001) |

Table.1. Switching Pattern for inverter in conventional DTC

Figure 4 shows the DTC block diagram which is included the speed and hysteresis controllers, flux and torque estimators and switching table. During the operation of the DTC switching system, the reference electromagnetic signal is produced by the PI speed controller using the error between the reference and the measured speeds. The errors between the desired and actual extent flux linkage and electromagnetic torque are forwarded to the hysteresis controller to produce the command based on the switching pattern table for the inverter [11,30].

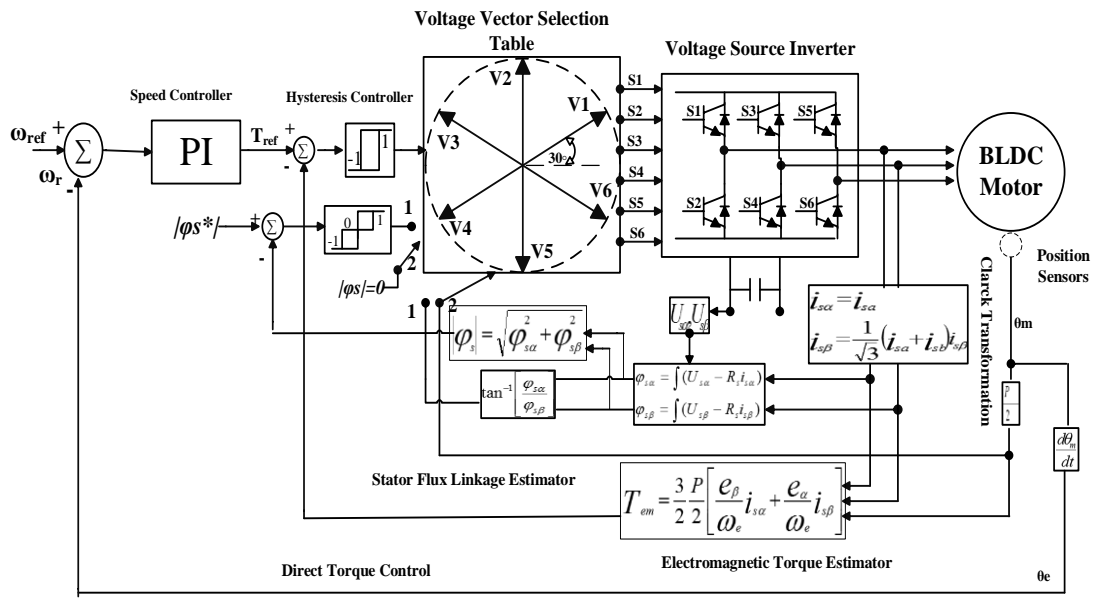

Figure. 4. General configuration of DTC with BLDC motor [18]

**The modified regenerative braking DTC**

In the motoring mode of BLDC motor stator flux vector moves ahead of the rotor flux. To produce inverse torque, the rotor and stator flux must be rotated in the opposite directions. As a result, the suitable voltage vector should be sent to the inverter in a way that the stator flux follows the rotor flux to return power to the battery. In conventional DTC for the switching pattern, there is no state to create inverse torque. Hence, the zeros vectors are selected when sharp drops when torque is required. Consequently, removing reversibility brings the motor speed down and makes dynamic response slower [32]. Selecting zeros vectors instead of vectors causes regeneration energy and rapid braking. In addition, reversibility brings the motor speed down and makes dynamic response slower [32].

In each sextet sector, a couple of active voltage vectors are determined to control flux-linkage magnitude. The exerted arrangements are shown in Figure 5.

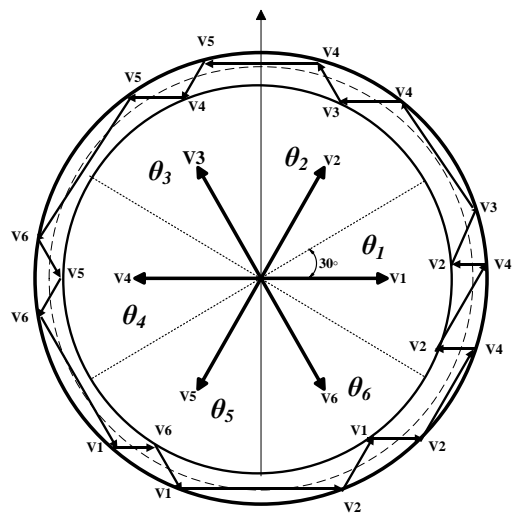

Figure. 5. Voltage vectors and their role

In this paper, the switching pattern has been modified to improve the conventional DTC method and regenerate energy from the braking system. Table 2 and Figure 6 show the vectors for different cases to increase or decrease the electromagnetic torque and flux and modify the results. Using the updating switching pattern, the negative torque is applied to the motor and the produced electrical energy which is transferred through the inverse diodes to charge the storage devices. In this method, the estimated flux error is applied to the two levels hysteresis controller. Also, a three levels hysteresis controller has been utilized to control the torque as:

$$Hystersis\ Output = \begin{cases} 1 & Increase\ the\ torque \\ 0 & No\ change \\ -1 & Decrease\ the\ Torque \end{cases}$$

By considering $k$ sector for the stator flux, the voltage vectors $V_{s,k+1}$ and inverse voltage vector $V_{s,k-2}$ are used to increase and decrease and the torque and the flux respectively. Also, two zero voltage vectors are applied to reach the unchanged flux and the low torque reduction. The modified switching pattern has been presented in table 3.

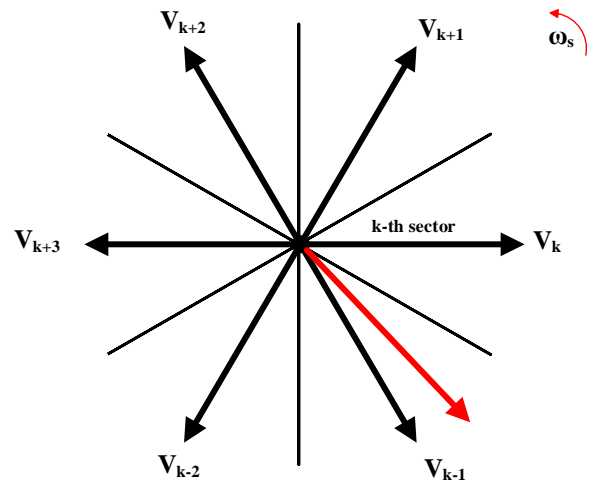

Figure. 6. The effect of selecting a voltage vector if the flux is in the k sector

| VOLTAGE VECTOR | Increase | Decrease |
|---|---|---|
| Flux | $V_k$, $V_{k+1}$, $V_{k-1}$ | $V_{k+3}$, $V_{k+2}$, $V_{k-2}$ |
| Torque | $V_{k+2}$, $V_{k+1}$ | $V_{k-2}$, $V_{k-2}$ |

Table. 2. Select vectors based on increasing or decreasing flux and torque

| $F_{st}$ | $T_{st}$ | Sector | | | | | |
|---|---|---|---|---|---|---|---|
| | | Ө1 | Ө2 | Ө3 | Ө4 | Ө5 | Ө6 |
| TI | FI | V2(001001) | V3(011000) | V4(010010) | V5(000110) | V6(100100) | V1(100001) |
| | FD | V3(011000) | V4(010010) | V5(000110) | V6(100100) | V1(100001) | V2(001001) |
| Tt | FI | V0(010101) | V7(101010) | V0(010101) | V7(101010) | V0(010101) | V7(101010) |
| | FD | V0(010101) | V7(101010) | V0(010101) | V7(101010) | V0(010101) | V7(101010) |
| TD | FI | V6(100100) | V1(100001) | V2(001001) | V3(011000) | V4(010010) | V5(000110) |
| | FD | V5(000110) | V6(100100) | V1(100001) | V2(001001) | V6(100100) | V4(010010) |

Table.3. Proposed switching table

**Model reference adaptive control**

Since the conventional PI controller is not robust in presence of parameters uncertainty and the disturbance signal [33-34], a MRAS controller is used to overcome these problems and guarantee the zero-tracking error. The block diagram of MRAS has been shown in Figure 7 [3].

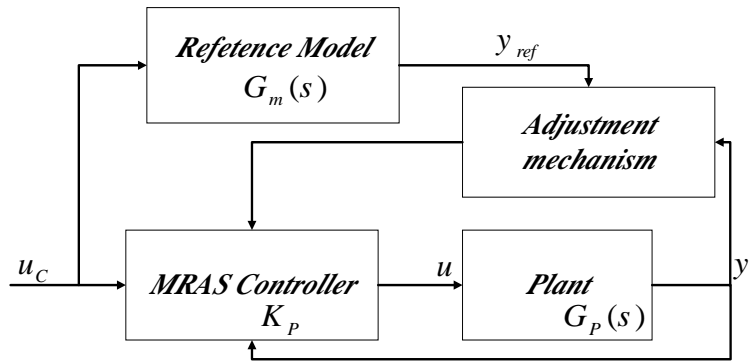

Figure.7. General form of MRAS

$G_m(s)$ is the model reference transfer function which is equal to $\bar{k}_p G_p(s)$. $G_p(s)$ is equivalent transfer function of the system and $k_p$ is the proportional adaptive controller parameter that controls the system in a way that the desired output signals track the reference signals. $u$ is the control signal which is equal to $k_p u_c$. The adaptive controller uses the error signal to build the desired command. The main objectives of the adaptive controller are to guarantee the stability of the system in presence of uncertainty and disturbances and zero tracking error. The tuning mechanism is known as MIT rule and the algorithm can be derived by control error:

$$e = y - y_m = L^{-1}\left(k_p G_p(s)U(s) - \bar{k}_p G_p(s)U(s)\right) \tag{10}$$

The cost function that should be minimized to adjust the adaptation gain will be as:

$$J(k_p) = \frac{1}{2}e^2(k_p) \tag{11}$$

It is rescannable to use the negative gradient of the J to adjust the adaptation parameter. The tracking error is computed as:

$$\begin{aligned}
e = y - y_m &= L^{-1}\left(G_p(s)k_p U(s) - G_p(s)\bar{k}_p U(s)\right) \\
&= L^{-1}\left(G_p(s)U(s)(k_p - \bar{k}_p)\right) \\
&= L^{-1}\left(\frac{G_p(s)U(s)\bar{k}_p(k_p - \bar{k}_p)}{\bar{k}_p}\right) = \frac{(k_p - \bar{k}_p)}{\bar{k}_p}L^{-1}\left(\bar{k}_p G_p(s)U(s)\right) \\
&= \frac{(k_p - \bar{k}_p)}{\bar{k}_p}L^{-1}(G_m(s)U(s)) = \frac{(k_p - \bar{k}_p)}{\bar{k}_p}L^{-1}(Y_m(s)) = \frac{(k_p - \bar{k}_p)}{\bar{k}_p}y_m
\end{aligned} \tag{12}$$

Also from the equation (11):

$$\begin{aligned}
J(k_p) &= \frac{1}{2}e^2(k_p) \\
\frac{dk_p}{dt} &= -\gamma \frac{\partial J}{\partial k_p} = -\gamma e \frac{\partial e}{\partial k_p}
\end{aligned} \tag{13}$$

Where $\frac{\partial e}{\partial k_p}$ is the sensitivity derivative of the system and $\gamma$ is the adaptation gain [35].

Using the Equations (12) and (13) the MIT adaptation rule will be:

$$\frac{dk_p}{dt} = -\gamma e \frac{\partial e}{\partial k_p} = -\gamma e y_m \tag{14}$$

**Simulation results and discussion**

Simulations have been done in MATLAB/Simulink to prove the efficiency of the proposed method. To illustrate the performance of the proposed method different scenarios have been considered. The system transfer function is given as:

$$Gp = \frac{b_0}{s^2 + bs + a} \quad (15)$$

The reference model is chosen to be:

$$Gm = \frac{b_m}{s^2 + bs + a} \quad (16)$$

where $a = R/L + B/J$, $b = RB/LJ$

## 6.1 Conventional DTC

A driving cycle which is similar to the standard ECE-driving-cycle, has been applied to the motor.

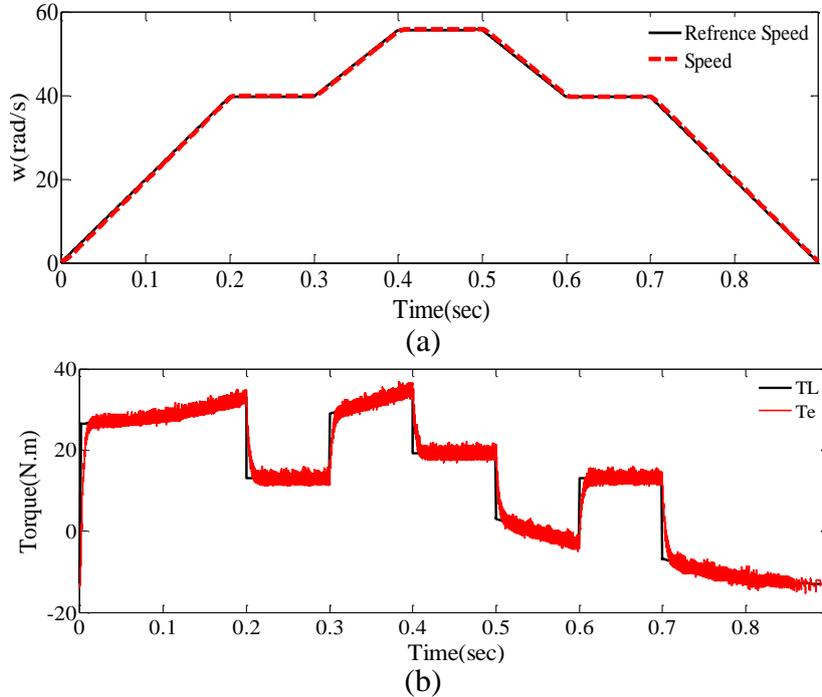

Figure. 8. (a) Speed tracking and demanded speed, (b) Torque tracking of BLDC motor in common DTC

In order to demonstrate the proper functioning of the DTC in BLDC motor, tracking torque and speed have been monitored. The speed and torque control of the conventional DTC based on the measured standard cycle and load torque are shown in Figure 8. As it can be seen, low torque ripple and fast dynamic response show the good performance of this method. Also, the output signals track the desired reference signals. The main difference between the conventional DTC and the proposed modified DTC is the State of Charge (SOC) of the systems which will be explained in the next sections.

## 6.2 Modified DTC

Figure 9 shows the speed and torque signals resulted by evaluating the modified DTC method. The driving cycle with a braking state and a negative acceleration, provides a situation that can be used to recover the kinetic energy and charge the batteries. The SOC of the battery in the regenerative braking mode has been shown in Figure 9 (c). As it can be seen from the result, the SOC of the battery has been improved. The faster speed reduction results in the significant drop in the SOC.

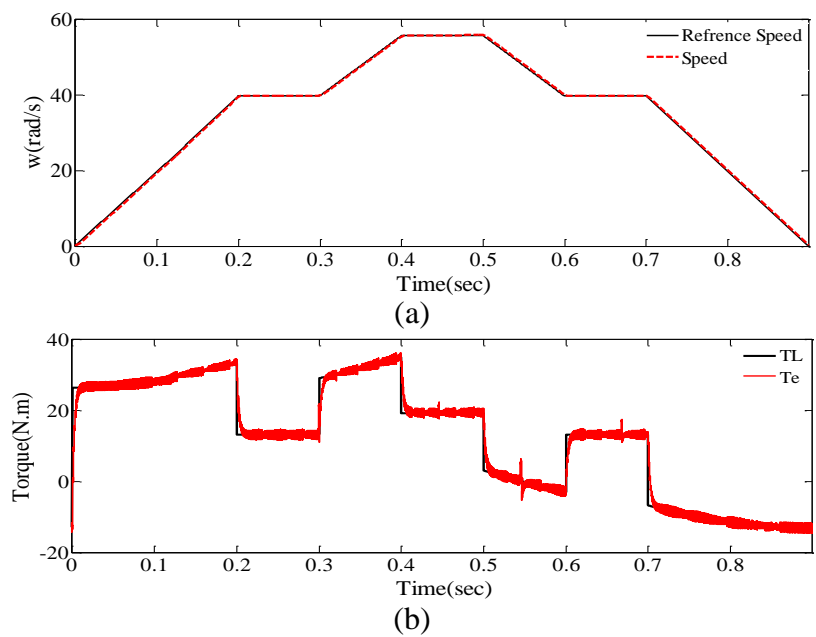

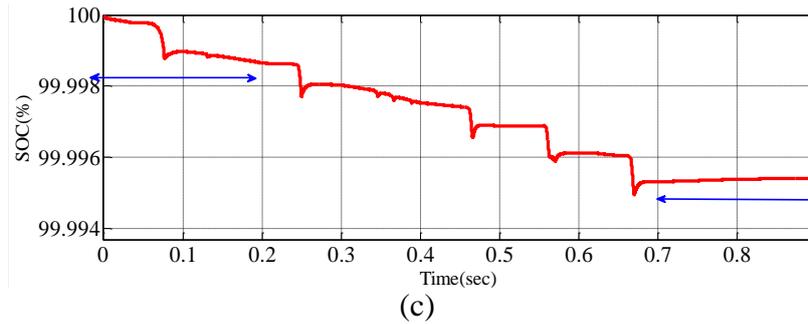

(c)

Figure. 9. (a) Speed tracking and references speed, (b) Actual Torque tacking with desired torque, (c) SOC in regeneration mode of DTC

## 6.3 SOC of the DTC and the modified DTC

In this scenario, the SOC of the conventional DTC and modified DTC have been compared in Figure 10. As it can be seen, in the same circumstances the modified DTC method has regenerated the power at the end of the cycle and has charged the batteries. However, in the conventional DTC method the SOC has decreased at the end of the cycle. By employing the modified DTC, the kinetic energy is converted to the electric energy when the acceleration is negative.

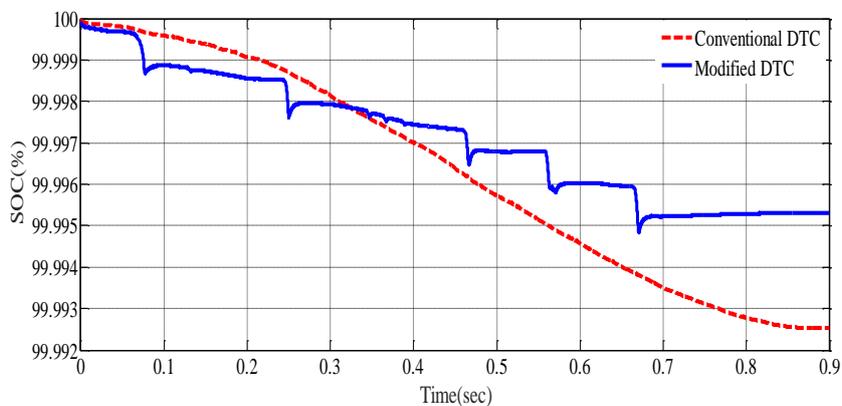

Figure. 10. SOC for battery with and without regenerative braking

## 6.4 MRAS controller

In the last scenario, an adaptive controller has been designed to control the system and improve the performance of the system. By applying the adaptive controller, the system will have a good performance to track the reference signals and also can help the system to damp the torque ripples as shown in Figure 11. By comparing the results of the modified DTC and the conventional DTC, the DTC method has less tracking error than the modified DTC, while the modified DTC removes the torque and speed ripples. To have a simple and reliable algorithm, the adaptive controller is a good option to satisfy the system's performance without changing the parameters. Based on the simulated results, not only the speed and the torque are tracked well, but also the torque ripple is significantly reduced in comparison with PI controller.

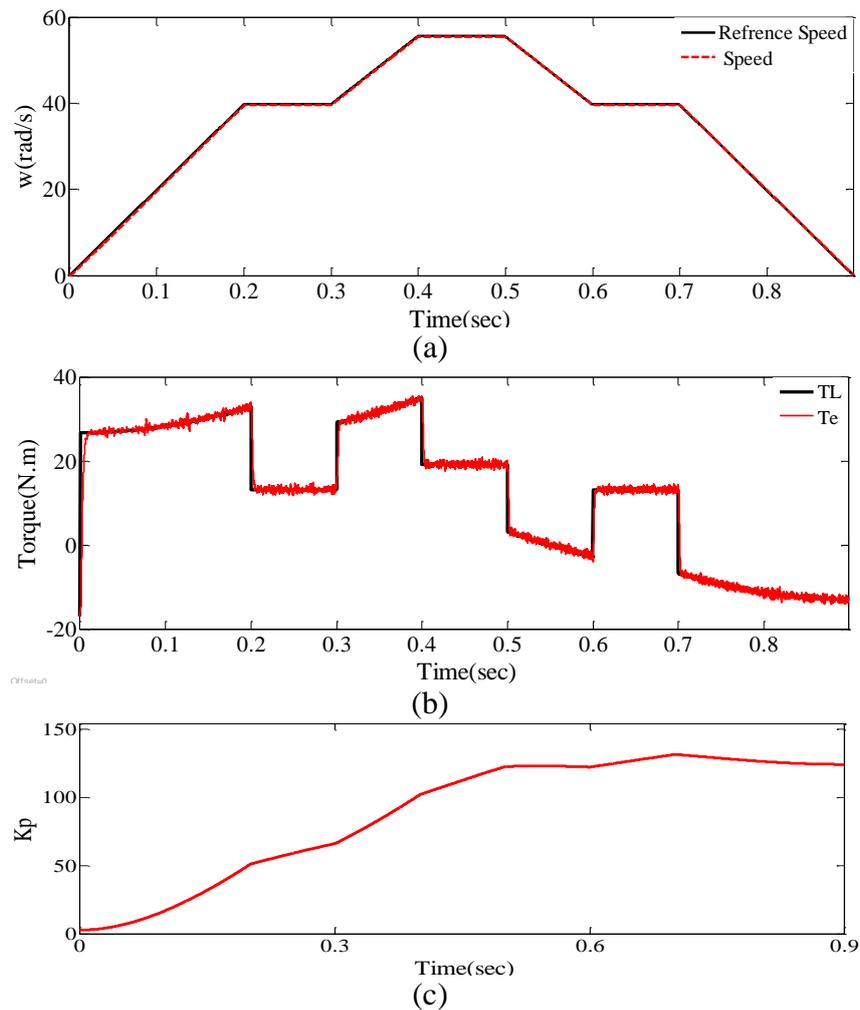

Figure. 11. (a) Speed tracking, (b) Torque, (c) Adaptive proportional gain

Also, adaptability is another important point to prove its efficiency, therefore a new reference signal is applied to monitor the speed tracking. As shown in Figure 12, PI controller cannot follow the new reference signal, whilst the proposed adaptive controller has a good performance without any changes in parameters. Also, comparing the SOCs clearly shows that battery energy usage is the lowest in this state.

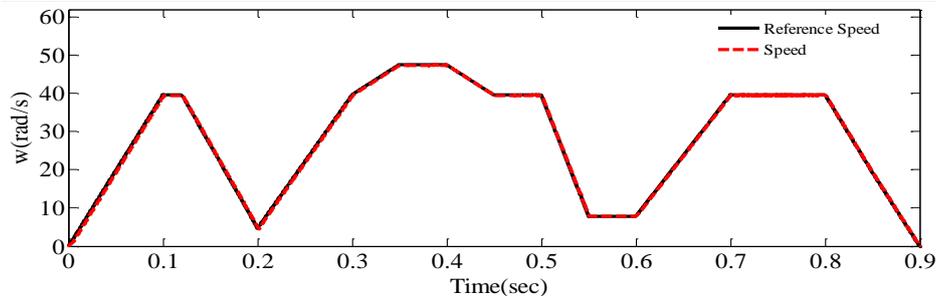

(a)

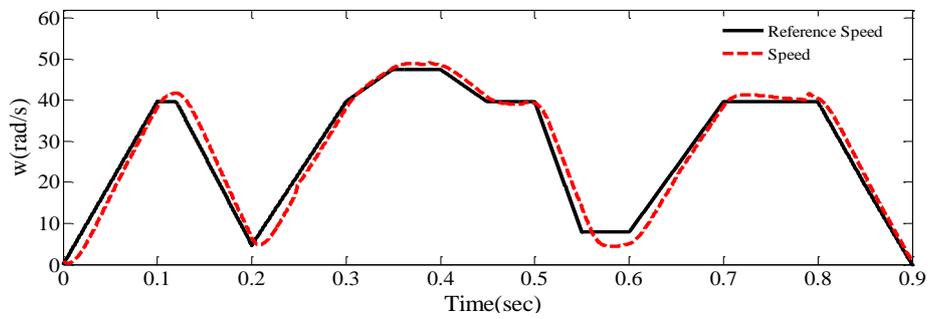

(b)

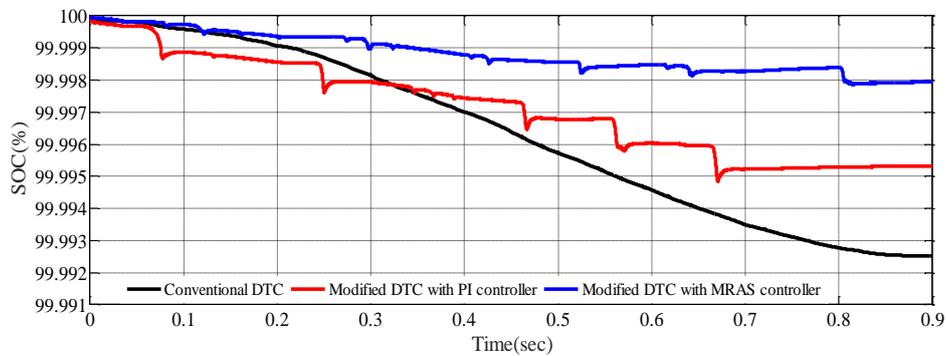

Figure. 12. (a) Speed tracking of MRAS, (b) Speed tracking of PI, (c) SOCs

**Conclusion**

In this study, a new approach is presented to regenerate electrical energy from the kinetic energy of EVs and bringing it back to the batteries in BLDC motor. Driving cycle experiences ups and downs with positive, negative and zero accelerations. These fluctuations lead to variations in the load torque. Consequently, the simulations are brought closer to the actual situation. In the proposed method, a new switching pattern has been implemented tothe system to generate the electrical energy without mechanical changes. The simulation results show that the new switching pattern improves the speed and torque tracking signals and the torque ripples. Comparing the SOC trends of conventional and modified DTC exhibits the good performance of the proposed algorithm and improvement in the energy return to batteries. In order to improve the tracking error and the torque ripples, an adaptive controller is designed. The results show the good performance of the designed controller in different speed reference signals.